\documentstyle[prl,aps,epsf,floats]{revtex}  

\begin{document}

\twocolumn[
\hsize\textwidth\columnwidth\hsize
\csname@twocolumnfalse\endcsname

\title{Measuring $V_{ub}$ in Inclusive $B$ Decays to Charm}
\author{Adam F. Falk and Alexey A. Petrov\\[4pt]}
\address{\tighten{\it
Department of Physics and Astronomy, 
The Johns Hopkins University,\\
3400 North Charles Street, 
Baltimore, Maryland 21218 USA}\\[4pt]
JHU--TIPAC--99003, hep-ph/9903518}

\maketitle

\tighten{
\begin{abstract}

We propose a novel method of measuring the CKM matrix element
$V_{ub}$, via inclusive $B$ decays to ``wrong sign'' charm. 
This process is mediated by the quark level decay $b\to u\bar
cs'$.  When normalized to the inclusive semileptonic decay
rate, the theoretical expression is very well behaved, with
small uncertainties from unknown quark masses and no
leading renormalon ambiguity.  We compute the decay
rate for this process, including the leading perturbative and
nonperturbative corrections.

\end{abstract}
}
\vspace{0.2in}

]\narrowtext

\newpage

The accurate measurement of the Cabibbo-Kobayashi-Maskawa
parameter $V_{ub}$ remains an important outstanding problem in
$B$ physics.  The magnitude of $V_{ub}$ corresponds to one of
the sides of the Unitarity Triangle, the angles of which will be
constrained or extracted from experiments to be performed soon
by the CDF, CLEO-III, BaBar and BELLE collaborations. 
Measuring $|V_{ub}|$ with precision would provide a vital
complementary check on the adequacy and consistency of the CKM
framework for the physics of flavor in the Standard
Model~\cite{review}.

The methods which are currently available for probing
$V_{ub}$ are unfortunately plagued by dependence on
phenomenological models whose uncertainties are difficult to
quantify reliably.  As a result, despite heroic
experimental efforts present constraints on this parameter
are unacceptably weak.  The analyses which have been used fall
into two classes, inclusive decays of the form $B\to
X_ul\nu$, and exclusive transitions such as
$B\to(\pi,\rho)l\nu$. 

The inclusive decay rate has the advantage that it can
be predicted in the form of a systematic expansion in powers of
$1/m_b$~\cite{ope}. However, the measurement of this process is
complicated by the overwhelming background of inclusive
$B\to X_cl\nu$ transitions.  At present, this background can
be suppressed only by confining oneself to kinematic
regions in which only charmless final states can contribute,
such as $E_l>2.3\,$GeV or $M(X)<1.9\,$GeV.  Unfortunately,
the same operator product expansion techniques which allow
one to calculate reliably the total inclusive rate break
down when the phase space is restricted in this way. 
Phenomenological models must then be used to reconstruct the
rate in the unobserved kinematic regions, and the model
independence of the analysis is lost~\cite{endpoint,massspec}.

Exclusive transitions such as $B\to(\pi,\rho)l\nu$ are easier
to study experimentally.  On the other hand theoretical
predictions of exclusive decay channels are polluted by our
ignorance of the physics of quark hadronization. Various
approaches to this problem have been proposed (for example,
heavy quark symmetry, chiral expansions in the soft pion limit,
dispersion relations, QCD sum rules, and lattice calculations),
but in many of these cases significant model dependence
remains.  It is therefore desirable to continue to look for a
model independent method for the extraction of $V_{ub}$.

We propose that $V_{ub}$ may be extracted from the inclusive
nonleptonic transition $b\to u\bar cs'$, where
$s'=s\cos\theta_C-d\sin\theta_C$ is the flavor eigenstate
which couples to $c$ and we take $m_s=m_d=0$.  Here
$\theta_C$ is the Cabibbo mixing angle, with
$\sin\theta_C\simeq0.2205$.  This process stands out for its
theoretical simplicity.  Because the transition involves
four distinct quark flavors, only diagrams proportional to
$V_{ub}$ enter the theoretical expression, with no
complications from penguins or strong rescattering
contributions.  (For similar reasons, it also has been proposed
recently that $V_{ub}$ may be extracted from {\it exclusive\/}
processes mediated by this quark transition~\cite{GrLe99}.) 
Moreover, as we show below, the ratio of this rate to that for
$b\to cl\nu$ receives well controlled radiative corrections,
with no leading infrared renormalon ambiguity.  We will compute
the largest nonperturbative corrections, which are
spectator-dependent and arise at order $1/m_b^3$.  We warn the
reader that our analysis will rely implicitly on the use of
parton-hadron duality.  This assumption is common to all
extractions of $V_{ub}$ from inclusive $B$ decays, and while it
is not unreasonable to expect it to hold in this case, it has
not been proven rigorously to do so.

Is such an analysis feasible experimentally?  What is
required is to find ``wrong-sign'' charmed states in
flavor-tagged $B$ decays.  This is certainly a challenging
measurement. However, some of the necessary techniques already
exist, as can be seen in Ref.~\cite{Mar99}.  In addition,
the combination of moving pairs of $B$ mesons and the excellent
vertexing capabilities of the the BaBar and BELLE detectors
should certainly improve the prospects for the measurement of
this transition~\cite{Kino99}.  Of course, the largest source of
``wrong-sign'' charm is from decays mediated by
the transition $b\to c\bar cs'$, which are enhanced roughly by a
factor of 100 over those which we propose to study.  Whether
such final states can be rejected at this level is an
experimental question which we are certainly not equipped to
answer.  However, we note that while numerically the problem
would seem to be similar to the quite intractable one of digging
$b\to ul\nu$ out of the $b\to cl\nu$ background, it is actually
quite different in its details.  In particular, in the case of
nonleptonic decays there is no missing neutrino, and perhaps
the enormous samples of $b\to c\bar cs'$ decays which will be
available to BaBar, BELLE and CLEO-III will allow one to
understand the kinematic features of those decays well enough
that they may be subtracted reliably.  In any case, our purpose
in this Letter is simply to argue that the theoretical
situation is so attractive that the feasibility of the
experiment is worth investigating.

We begin with the transition $b\to\bar cX$ at lowest
order.  The decay is mediated by the local $\Delta B=1$
Lagrangian,
\begin{eqnarray}\label{lagr}
  {\cal L} =&& \frac{4G_F}{\sqrt{2}} \,V_{ub} \,\Bigl[
  c_1(\mu) \bar u_\alpha \gamma_\mu P_Lb_\alpha 
  \,\bar s'_\beta \gamma^\mu P_L c_\beta \nonumber\\
  &&\qquad\mbox{}+c_2(\mu)\bar u_\alpha \gamma_\mu P_L b_\beta 
  \,\bar s'_\beta \gamma^\mu P_L c_\alpha\Bigr]+{\rm h.c.}
  \nonumber\\
  =&& \frac{4G_F}{\sqrt{2}} \,V_{ub} \,\Bigl[
  \Big(c_1(\mu)+{1\over N_c}\,c_2(\mu)\Big) \bar u
  \gamma_\mu P_Lb 
  \,\bar s' \gamma^\mu P_L c \nonumber\\
  &&\qquad\mbox{}+2c_2(\mu)\,\bar u\gamma_\mu
   T^aP_L b\,\bar s' \gamma^\mu T^a P_L c\Bigr]+{\rm h.c.},
\end{eqnarray}
where $P_L$ is the left-handed projector and $N_c=3$ is the
number of colors.  At tree level, $c_1=1$ and $c_2=0$; to
leading logarithmic order, taking $\alpha_s(m_Z)=0.117$, the
Wilson coefficients at the scale $m_b$ are given by
$c_1(m_b)\simeq1.11$ and
$c_2(m_b)\simeq-0.25$~\cite{Leff}.

The inclusive decay rate in the channel of interest may be
written as the imaginary part of the matrix element of the
forward transition operator ${\cal T}$,
\begin{eqnarray}\label{rate1}
  \Gamma (b \to \bar c X) &=&
  \langle B|\,{\cal T} (b\to\bar cX\to b)\,|B\rangle\\
  &=&\frac{1}{m_b} \langle B |\,{\rm Im}\,\Big\{
  i \int {\rm d}^4x\, T\,[
  {\cal L}^\dagger(x),{\cal L}(0)]\Big\}| B \rangle\,.
  \nonumber
\end{eqnarray}
At lowest order, for which the quark level process is $b\to
u\bar cs'$, we parameterize the $\bar cs'$ loop by
\begin{equation}
  \Pi^{\bar c s'}_{\mu \nu} (q) = \frac{N_c}{8 \pi} 
  \left[ A(q^2) \left(q_\mu q_\nu -q^2 g_{\mu \nu}\right) +
  B(q^2) q_\mu q_\nu\right],
\end{equation}
where $q^\mu$ is the
momentum carried by the $\bar c s'$ pair.  Then the decay rate
may be written as~\cite{Vol95}
\begin{eqnarray}\label{rate2}
  \Gamma (b \to u \bar c s') = 3 N_c \Gamma_0  |V_{ub}|^2
  x_c^4\int_{1}^{1/x_c}
  {\rm d}z \left(1/x_c - z\right)^2\nonumber\\ 
  \times
  \left[ A(z) \left(2z+1/x_c\right)
  +B(z)/x_c \right],
\end{eqnarray}
with $x_c=m_c^2/m_b^2$ and $\Gamma_0=G_F^2 m_b^5/192
\pi^3$.  At tree level, with $q^2=zm_c^2$, the form factors
$A_0(z)$ and  $B_0(z)$ are
\begin{eqnarray}
  A_0(z) &=& \chi\, a_0(z)= 
  \textstyle{2\over3}\chi \left( 1 - 1/z \right)^2
  \left(1 + 1/2z \right), \nonumber \\
  B_0(z) &=&\chi\, b_0(z)= \chi\left( 1 - 1/z \right)^2/z
\end{eqnarray}
with $\chi=(c_1+c_2/N_c)^2+2 c_2^2/N_c\simeq1.09$. 
The two terms in $\chi$ come respectively from the squares of
the color singlet and color octet operators in ${\cal L}$. 
Neglecting the octet operator would induce only a four
percent error in $\chi$.  Integrating over $z$ and normalizing
the result to the semileptonic decay rate $\Gamma(b\to
cl\nu)=\Gamma_0 |V_{cb}|^2f(x_c)$,
where
\begin{equation}
  f(x_c)=1-8x_c+8x_c^3-x_c^4-12x_c^2\ln x_c,
\end{equation}
yields a very simple expression at leading order,
\begin{eqnarray}\label{ratio0}
  R&=&
  \Gamma(b\to\bar cX)/N_c\Gamma(b\to cl\nu)\nonumber\\
  &=&\chi\left|V_{ub}/V_{cb}\right|^2 \Big\{1 + 
  {\cal O}(\alpha_s, 1/m_b^2) \Big\}.
\end{eqnarray}
Note that at this order the ratio $R$ is independent of the
heavy quark masses.  With no corrections included, the only
distinction between these two processes is in the
kinematics of the final state products. However, the phase
space function $f(x_c)$ turns out to be the same in the two
cases and cancels in the ratio.  The only dependence on $x_c$
will be in the higher order corrections to $R$.  Hence the
effect of uncertainties and ambiguities in the definition and
extraction of $m_c$, $m_b$ and $x_c$ will be much smaller in
$R$ than in the individual rates.

We now turn to the effects of perturbative QCD.  The
radiative correction to $\Gamma(b\to cl\nu)$ comes from the
finite renormalization of the $\bar c_L\gamma^\mu b_L$ current,
plus gluon bremsstrahlung.  It has been computed analytically
by Nir~\cite{Nir89}, and the result can be put into the form
$\Gamma(b\to
cl\nu)=\Gamma_0|V_{cb}|^2f(x_c)[1+a_{bc}(x_c)\,\alpha_s/\pi]$. 
The radiative correction to $\Gamma (b\to u\bar cs')$ is more
complicated, and we will only consider the corrections
to the color singlet transition.  In this case we can neglect
gluons which propagate from the $bu$ line to the $\bar cs'$
line, simplifying the calculation enormously.  As noted above,
at tree level the color octet transition only contributes
four percent to the total decay rate, so the error in
omitting the radiative correction to this tiny contribution
should be small.  Of course, this is a phenomenological
approximation, which is not formally consistent in terms of
an expansion in leading logarithms.  Here we sacrifice formal
consistency in favor of identifying and keeping the terms which
are numerically the largest.  A more complete analysis is
presented in Ref.~\cite{Chay99}.

\begin{figure}
\centerline{
\epsfysize 2.0in
\epsfbox{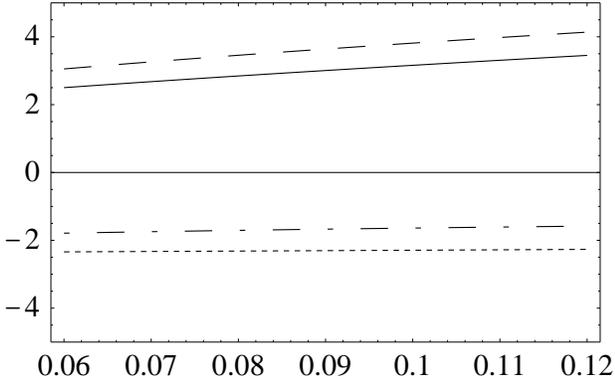}}
\vskip0.5cm
\caption{Radiative corrections to $R$ as a function of
$x_c$.  We display curves for $a_{\bar cs'}$
(dashed), $a_{bu}$ (dotted), $a_{bc}$
(dotted-dashed), and the total $g(x_c)$ (solid).}
\label{radiative}
\end{figure}

In this framework, then, there remain two classes of
radiative corrections to $\Gamma (b\to u\bar cs')$.  The
first, $a_{bu}$, comes from finite renormalization of the
$\bar u_L\gamma^\mu b_L$ current.  This correction, which can
be extracted from an analysis of $\Gamma(b\to
u\tau\nu)$ by Czarnecki, Je\.zabek and K\"uhn~\cite{CJK95},
also depends on $x_c$ through the phase space of the $\bar cs'$
pair.  The second contribution, $a_{\bar cs'}$, comes from the
radiative corrections to the $\bar cs'$ loop~\cite{Vol95,Re80}. 
The effect, in the case of the color singlet operators, is to
add corrections to the form factors in $\Pi_{\mu\nu}^{\bar
cs'}(q)$,
\begin{eqnarray} 
  A_1(z) &=& \chi a_0(z)
  {\frac{4\alpha_s}{3\pi}}\left[
  f_1(z)+{2z\over1+2z}\,f_2(z) \right], \nonumber
  \\ B_1(z) &=& \chi b_0(z)
  {\frac{4\alpha_s}{3\pi}}\Big[f_1(z)-1\Big],
\end{eqnarray}
where $f_1(z)$ and $f_2(z)$ are functions of the charm quark
mass which have been computed by Voloshin~\cite{Vol95}.  We
find a radiative correction to $R$ of the form
\begin{equation} \label{ratioalpha}
  R = \chi\left|V_{ub}/V_{cb}\right|^2  \Big\{
  1 + g(x_c)\,{\alpha_s\over\pi}+\dots\Big\},
\end{equation}
with $g(x_c)=a_{bu}(x_c) - a_{bc}(x_c) + a_{\bar cs'}(x_c)$. 
The function $g(x_c)$, along with each individual term, is
plotted in Fig.~\ref{radiative} for $4.5\,{\rm GeV}\le m_b\le
4.9\,{\rm GeV}$, which via the relation
$m_b-m_c=m_B-m_D=3.34\,{\rm GeV}$ corresponds to the range
$0.06\le x_c\le 0.12$.  For the ``central value'' $x_c=0.09$,
and with $\alpha_s(m_b)=0.22$, one has
$g(x_c)\alpha_s/\pi=0.21$.  We see that the one loop radiative
correction is fairly large.

In principle, we could also include corrections of order
$\alpha_s^2\beta_0$.  Such terms, and all those of the form
$\alpha_s^n\beta_0^{n-1}$, play an important role in the
resolution of infrared renormalon ambiguities associated with
dependence on the pole mass $m_b$~\cite{renormalons}.  In
particular, if $m_b$ (or equivalently, the Heavy Quark
Effective Theory (HQET) parameter $\bar\Lambda$) appears in an
expression at leading order, then the associated perturbative
series has a renormalon ambiguity at order $\Lambda_{\rm
QCD}/m_b$, and the term proportional to $\alpha_s^2\beta_0$
typically is large and must be included.  In the present case,
however, the leading dependence $m_b^5f(x_c)$ cancels in the
ratio $R$, and therefore the leading infrared renormalon
cancels as well.  As a result, there is no reason to expect two
loop corrections to be enhanced, even those proportional to
$\alpha_s^2\beta_0$.  It is another attractive feature of the
process $b\to u\bar cs'$ that, when normalized to $b\to
cl\nu$, its perturbative series has no leading renormalon and
is well behaved.

We now turn to the nonperturbative corrections to $R$. 
Because $R$ is independent of $m_b$ and $m_c$ at leading
order, the first corrections appear at order $1/m_b^2$.  They
are proportional to the HQET parameters $\lambda_1$ and
$\lambda_2$, which parameterize bound state effects (kinetic
energy and hyperfine interactions) of the $b$ quark in the
initial $B$ meson~\cite{lambda}.  These corrections were
calculated for the process $b\to c\tau\nu$ in
Ref.~\cite{Fa94}.  The results of that calculation apply to
$b\to u\bar cs'$ with the identifications $m_c\to0$ and
$m_\tau\to m_c$; they apply to
$b\to cl\nu$ in the limit $m_\tau\to0$.  However, inspection
of Eq.~(2.15) of Ref.~\cite{Fa94} reveals that the correction
is actually {\it symmetric\/} in $m_c$ and $m_\tau$, and 
hence the corrections at order $1/m_b^2$ cancel exactly in $R$.

The leading nonperturbative corrections arise, then, at order
$1/m_b^3$.  Two types of contribution appear at this order. 
First, there are additional bound state effects, which may be
written in terms of six new HQET parameters $\rho_1$,
$\rho_2$, and ${\cal T}_1,\ldots {\cal T}_4$~\cite{mcubed}.  We
have no reason to suppose that these contributions should be
enhanced over their usual size, typically at the percent level
in total rates.  On the contrary, we might expect somewhat of a
cancellation in $R$ as obtained for
$\lambda_1$ and $\lambda_2$.  We will not include such terms
here.  The second type of effect comes from annihilation
processes such as $b\bar u\to\bar cs'$.  These diagrams
contribute only to nonleptonic decays, hence only to the
numerator of $R$.  In addition, they are expected to be
enhanced over other terms at order $1/m_b^3$ by a phase space
factor of $16\pi^2$, corresponding to a two-body instead of
a three-body final state.  Finally, they contribute
differently to $B^-$ and $B^0$ decays, and can be probed by
comparing $R$ in the two cases.  (Similar processes are
responsible for the lifetime difference
$\tau(B^-)\neq\tau(B^0)$~\cite{VoSh85,BlSh93,NeSa97}.)

Annihilation diagrams are associated with four quark operators
of dimension six.  Performing the operator product expansion to
this order, we find a contribution to the transition
operator for $B^-$ and
$\overline B{}^0$ decays,
\begin{eqnarray}
  {\cal T}_{\rm sp}&&=
  {G_F^2m_b\over3\pi}\,|V_{ub}|^2\,(1-x_c)^2\Big\{
  N_c(c_1+c_2/N_c)^2\nonumber\\
  &&\qquad\times\left[(1+2x_c)O^u_{S-P}
  -(1+x_c/2)O^u_{V-A}\right]\nonumber\\
  &&\,\mbox{}+2c_2^2\left[(1+2x_c)T^u_{S-P}
  -(1+x_c/2)T^u_{V-A}\right]\nonumber\\
  &&\,\mbox{}+
  \sin^2\theta_C\,N_c(c_2+c_1/N_c)^2\nonumber\\
  &&\qquad\times\left[(1+2x_c)O^d_{S-P}
  -(1+x_c/2)O^d_{V-A}\right]\\
  &&\,\mbox{}+2\sin^2\theta_C\,c_1^2\left[(1+2x_c)T^d_{S-P}
  -(1+x_c/2)T^d_{V-A}\right]\Big\},\nonumber
\end{eqnarray}
where we define~\cite{NeSa97}
\begin{eqnarray}
  O^q_{V-A}&=&\bar b_L\gamma^\mu q_L\,\bar q_L\gamma_\mu b_L
  \,,\nonumber\\
  O^q_{S-P}&=&\bar b_Rq_L\,\bar q_L b_R
  \,,\nonumber\\
  T^q_{V-A}&=&\bar b_L\gamma^\mu T^aq_L\,
  \bar q_L\gamma_\mu T^ab_L
  \,,\nonumber\\
  T^q_{S-P}&=&\bar b_RT^aq_L\,\bar q_L T^ab_R\,.
\end{eqnarray}
The operators $O^q_{V-A}$ and $O^q_{S-P}$ are color
singlet, while $T^q_{V-A}$ and $T^q_{S-P}$ are color octet. 
Also, note that the terms mediating $\overline B{}^0$ decays
are suppressed by an additional factor of
$\sin^2\theta_C$, because annihilating the initial state
requires an operator with $q=d$ rather than
$q=s'$.

The contribution of ${\cal T}_{\rm sp}$ to $B$ decays is
obtained by evaluating the forward matrix elements $\langle
B_q|\,{\cal T}_{\rm sp}\,|B_q\rangle$.  The matrix elements
of the four quark operators depend on nonperturbative QCD
and have not been computed in a controlled way.  Although
models such as quenched lattice QCD and QCD sum rules
can yield estimates of their values, reliable information will
probably have to wait until full unquenched lattice analyses
are available.  It is convenient to parameterize the matrix
elements in terms of ``bag parameters'' $B_i$ and
$\epsilon_i$~\cite{NeSa97},
\begin{eqnarray}
  \langle B_q|\,O^q_{V-A}\,|B_q\rangle&=&
  \textstyle{1\over4}\,f_{B_q}^2m_{B_q}^2\,B_1\,,\nonumber\\
  \langle B_q|\,O^q_{S-P}\,|B_q\rangle&=&
  \textstyle{1\over4}\,f_{B_q}^2m_{B_q}^2\,B_2\,,\nonumber\\
  \langle B_q|\,T^q_{V-A}\,|B_q\rangle&=&
  \textstyle{1\over4}\,f_{B_q}^2m_{B_q}^2\,
  \epsilon_1\,,\nonumber\\
  \langle B_q|\,T^q_{S-P}\,|B_q\rangle&=&
  \textstyle{1\over4}\,f_{B_q}^2m_{B_q}^2\,
  \epsilon_2\,.
\end{eqnarray}
In the vacuum insertion ansatz, we have $B_1=B_2=1$ and
$\epsilon_1=\epsilon_2=0$, and only the color singlet operators
contribute to the decay.  In an expansion in powers of $1/N_c$,
one finds $B_i={\cal O}(1)$, while $\epsilon_i={\cal O}(1/N_c)$;
in this more general case, the color octet matrix elements are
suppressed but nonvanishing.  For example, one QCD sum rules
estimate gives $\epsilon_1\approx-0.15$ and
$\epsilon_2\approx0$~\cite{Ch95}.  In terms of these
parameters, the contributions to 
$R(B^-,\overline B{}^0)=\chi\left|V_{ub}/V_{cb}\right|^2 
[1 + g(x_c)\alpha_s/\pi +\delta_3(B^-,\overline B{}^0)]$ 
are given by 
\begin{eqnarray}\label{rate3}
  \delta_3(B^-)&=&{16 \pi^2 f_B^2(1-x_c)^2
  \over\chi N_cm_b^2f(x_c)}\,
  \Big\{N_c(c_1+c_2/N_c)^2\nonumber\\
  &&\qquad\qquad\qquad\times\left[(1+2x_c)B_2
  -(1+x_c/2)B_1\right]\nonumber\\
  &&\quad\mbox{}+2c_2^2\left[(1+2x_c)\epsilon_2
  -(1+x_c/2)\epsilon_1\right]\Big\},\nonumber\\
  \delta_3(\overline B{}^0)&=&{16 \pi^2 f_B^2(1-x_c)^2
  \over\chi N_cm_b^2f(x_c)}\,\sin^2\theta_c\,
  \Big\{N_c(c_2+c_1/N_c)^2\nonumber\\
  &&\qquad\qquad\qquad\times\left[(1+2x_c)B_2
  -(1+x_c/2)B_1\right]\nonumber\\
  &&\quad
  \mbox{}+2c_1^2\left[(1+2x_c)\epsilon_2
  -(1+x_c/2)\epsilon_1\right]\Big\},
\end{eqnarray}
With $x_c=0.09$ and $f_B=200\,$MeV, we find
\begin{eqnarray}
  &&\delta_3(B^-)=-0.46B_1+0.52B_2
  -0.018\epsilon_1+0.020\epsilon_2\,,\\
  &&\delta_3(\overline B{}^0)=-0.00032B_1+0.00037B_2
  -0.017\epsilon_1+0.020\epsilon_2\,.\nonumber
\end{eqnarray}
We see that the annihilation process is much larger in charged
than in neutral $B$ decays.  Using factorization, with $B_i=1$
and $\epsilon_i=0$, we have $(\delta_3(B^-),\delta_3(\overline
B{}^0))=(0.059,4.2\times10^{-5})$; taking instead the sum
rules estimate $\epsilon_2=-0.15$ gives
$(\delta_3(B^-),\delta_3(\overline
B{}^0))=(0.062,2.6\times10^{-3})$.  Since it depends dominantly
on the largely unconstrained octet matrix element
$\epsilon_i$, there is huge uncertainty in $\delta_3(\overline
B{}^0)$; on the other hand, for any reasonable values of the
bag parameters the annihilation contribution to $\overline
B{}^0$ decay is below the percent level and is safely
negligible.  By contrast, $\delta_3(B^-)$ is much larger, but
since it depends mostly on the singlet matrix elements $B_i$,
it has a smaller fractional uncertainty.

To summarize, we have computed the inclusive branching
fraction for ``wrong sign'' charm production in $B$ decays,
normalized to inclusive semileptonic decay.  The leading
perturbative correction $g(x_c)\alpha_s/\pi$ is analytically
calculable, independent of the spectator quark, and
approximately $20\%$.  The annihilation contribution $\delta_3$
is smaller but more uncertain; it enters at the $5\%$ level for
$B^-$ decays and is negligible for $\overline B{}^0$ decays. 
There is no leading renormalon ambiguity in $R$, and dependence
on quark masses appears only in the radiative and
nonperturbative corrections themselves.  As a result, this
process is an excellent place to extract $|V_{ub}|$ from a
theoretical point of view.  This measurement will be
challenging experimentally, but given the crucial role of
$V_{ub}$ in constraining the Unitary Triangle and the present
lack of a viable alternative, it certainly is worth pursuing.

\bigskip

After this letter was submitted, we became aware that the
attractiveness of this process for probing $V_{ub}$ has been
noted previously~\cite{previous}, although not discussed in any
detail.  

\bigskip

We are grateful to V.~Sharma and F.~W\"urthwein for
discussions concerning the experimental feasibilty of this
measurement, and for encouraging us to publish this result.  We
also thank M.~Neubert, J.~Rosner and A.~Vainshtein for helpful
conversations. Support was provided by the NSF under Grant
PHY--9404057 and National Young Investigator Award
PHY--9457916, and by the DOE under Outstanding Junior
Investigator Award DE--FG02--94ER40869.  A.F.~is supported by
the Alfred P.~Sloan Foundation and by the Research Corporation
as a Cottrell Scholar.

\tighten


\begin{references}

\bibitem{review} For a review, see {\sl The BaBar
Physics Book,} eds. P.F. Harrison and H.R. Quinn, SLAC--R--504,
1998.

\bibitem{ope} J. Chay, H. Georgi and B. Grinstein, Phys.
Lett. B {\bf 247}, 399 (1990); M. Voloshin and M. Shifman, Sov.
J. Nucl. Phys. {\bf 41}, 120 (1985); I.I. Bigi {\it et al.},
Phys. Lett. B {\bf 293}, 430 (1992), Erratum, {\it ibid}, {\bf
297}, 477 (1993). 

\bibitem{endpoint} M. Neubert, Phys. Rev. D {\bf 49}, 3392
(1994); A.F. Falk {\it et al.}, Phys. Rev. D {\bf 49}, 4553
(1994).

\bibitem{massspec} A.F. Falk, Z. Ligeti and M.B. Wise, Phys.
Lett. B {\bf 406}, 225 (1997); I.I. Bigi, R.D. Dikeman and N.G.
Uraltsev, Eur. Phys. J. C {\bf 4}, 453 (1998).

\bibitem{GrLe99} B. Grinstein and R. Lebed, hep-ph/9902369;
D.H. Evans, B. Grinstein and D.R. Nolte, hep-ph/9903480.

\bibitem{Mar99} S. Marka, presentation VB09.10 at the
April 1999 Meeting of the APS; M. Bishai {\it et al.} (CLEO
Collaboration), Phys. Rev. {\bf D57}, 3847 (1998).

\bibitem{Kino99} K.~Kinoshita, private communication.

\bibitem{Leff} G. Altarelli and L. Maiani, Phys. Lett. {\bf
52B}, 351 (1974); M.K. Gaillard and B.W. Lee, Phys. Rev. Lett.
{\bf 33}, 108 (1979); F.G. Gilman and M.B. Wise, Phys. Rev.
D {\bf 20}, 2392 (1979).

\bibitem{Vol95} M.B. Voloshin, Phys. Rev. D {\bf 51}, 3948
(1995).

\bibitem{Nir89} Y. Nir, Phys. Lett. B {\bf 221}, 184 (1989).
 
\bibitem{Chay99} J. Chay, A.F. Falk, M. Luke and A.A. Petrov,
hep-ph/9907363, to appear in Phys. Rev. {\bf D}.

\bibitem{CJK95} A. Czarnecki, M. Je\.zabek and J.H. K\"uhn,
Phys. Lett. B {\bf 346}, 335 (1995).

\bibitem{Re80} L.J. Reinders, H.R. Rubinstein, and S. Yazaki,
Phys. Lett. {\bf 97B}, 257 (1980); {\it ibid.} {\bf 103B}, 63
(1981).

\bibitem{renormalons} M. Beneke and V.M. Braun, Nucl. Phys.
{\bf B426}, 301 (1994); I.I. Bigi {\it et al.}, Phys. Rev.
D {\bf 50}, 2234 (1994); M. Luke, A.V. Manohar and M.J. Savage,
Phys. Rev. D {\bf 51}, 4924 (1995).

\bibitem{lambda} A.F. Falk and M. Neubert, Phys. Rev. D
{\bf 47}, 2965 (1993); A.V. Manohar and M.B. Wise, Phys. Rev.
D {\bf 49}, 1310 (1994);  I.I. Bigi {\it et al.}, Phys. Rev.
Lett. {\bf 71}, 496 (1993).

\bibitem{Fa94} A.F. Falk {\it et al.}, Phys. Lett. B {\bf 326},
145 (1994); L. Koyrakh, Phys. Rev. D {\bf 49}, 3379 (1994).

\bibitem{mcubed} T. Mannel, Nucl. Phys. {\bf B413}, 396 (1994);
I.I. Bigi {\it et al.}, Phys. Rev. D {\bf 52}, 196 (1995); M.
Gremm and A. Kapustin, Phys. Rev. D {\bf 55}, 6924 (1997).

\bibitem{VoSh85} M.B. Voloshin and M.A. Shifman, 
Sov. J. Nucl. Phys. {\bf 41}, 120 (1985).

\bibitem{BlSh93} B. Blok and M.A. Shifman, Nucl. Phys. {\bf 
B389}, 534 (1993).

\bibitem{NeSa97} M. Neubert and C. Sachrajda, Nucl. Phys.
{\bf B483}, 339 (1997).

\bibitem{Ch95} V. Chernyak, Nucl. Phys. {\bf B457}, 96 (1995).

\bibitem{previous} M. Beneke, G. Buchalla and I. Dunietz, Phys.
Lett. B {\bf 393}, 132 (1997); I. Dunietz and J.L. Rosner,
CERN-TH-5899-90 (1990) (unpublished).

\end{references}
\end{document}